# E-commerce in Hungary: A Market Analysis

SZABOLCS NAGY, PH.D.
ASSOCIATE PROFESSOR

UNIVERSITY OF MISKOLC
email:marvel@uni-miskolc.hu

*SUMMARY*

*E-commerce is on the rise in Hungary, with significantly growing numbers of customers shopping online. This paper aims to identify the direct and indirect drivers of the double-digit growth rate, including the related macroeconomic indicators and the Digital Economy and Society Index (DESI). Moreover, this study provides a deep insight into industry trends and outlooks, including high industry concentration and top industrial players. It also draws the profile of the typical online shopper and the dominant characteristics of online purchases. Development of e-commerce is robust, but there is still plenty of potential for growth and progress in Hungary.*
*Journal of Economic Literature (JEL) codes: M31, L81*
*DOI: http://dx.doi.org/10.18096/TMP.2016.03.03*

## INTRODUCTION

With its 18% growth rate, e-commerce is the engine of growth for the whole retail sector in Hungary. E-commerce growth rate has been fueled by improving economic conditions and by the increasing number of Hungarians shopping for goods and services online. This rapid growth has turned e-commerce into an attractive field of research in academia. However, to the best of my knowledge, a comprehensive, thorough analysis of the e-commerce industry in Hungary is not available in the literature. This paper aims to eliminate this deficiency by analyzing the direct and indirect drivers of e-commerce growth and giving deeper insight into industry trends and outlooks.

It is inevitable that e-commerce plays a vital role in developing the economy. Anvari & Norouzi (2016) found that E-commerce and R&D are significant influencers of GDP (Gross Domestic Product) per capita based on purchasing power parity. However, e-commerce has a stronger development-enhancing impact compared to R&D. It is also confirmed in the literature that the Internet is the strongest growth engine of global trade nowadays (Gabrielsson & Gabrielsson 2011).

Competitiveness and complexity in the retail industry are growing because of rapid technological changes and diffusion (Pantano et al. 2017). Falk and Hagsten (2015, p. 10) stated that "while engagement in e-sales is still not widespread, it is more frequently used by large firms, high-productivity firms, and firms with international experience. The proportion of e-sales continues to grow over time from its low initial level. An increase in e-sales by one percentage point raises labour productivity growth by 0.3 percentage points over a two-year period". Choshin & Ghaffari (2017) added that "e-commerce is regarded as an appropriate strategy for marketing, selling and integrating online services which can play a significant role in identifying, obtaining and maintaining customers." Moreover, to some extent, online commerce has a positive impact (performance implication) on traditional trade (physical store retailers) through more intense competition (Johansson & Kask 2017).

It is almost generally accepted that the Internet plays an increasingly important role in our lives all over the world, creating chances for e-business and a-commerce (Apăvăloaie 2014). Chaparro-Peláez et al. (2016) concluded that there is no single motivation to drive consumers to shop online, but there are some combinations of motivations. Therefore, online retailers must satisfy these motivations in their product offering to guarantee full customer satisfaction. They also reconfirmed that convenience was still the main motivation to buy online; without convenience, no online purchase experience is possible.

Statista (2016), the biggest portal for statistics, defined e-commerce as follows: "The e-Commerce market encompasses the sale of physical goods via a digital channel to a private end user (B2C). Incorporated in this definition are purchases via desktop computer (including notebooks and laptops) as well as purchases via mobile devices such as smartphones and tablets. The following are not included in the e-Commerce market: digitally distributed services (eServices), digital media downloads or streams, digitally distributed goods in B2B markets nor digital purchase or resale of used, defective or repaired goods (re-Commerce and C2C). ". When it comes to analyzing the e-commerce market, I applied this above definition of e-commerce.

# BACKGROUND: THE DRIVERS OF THE GROWTH OF E-COMMERCE

Drivers of e-commerce growth in Hungary can be divided into two groups: direct and indirect drivers. The development level of the digital economy, especially internet and device use, is the key direct driver, while net monthly earnings, inflation and household consumption have an indirect influence on e-commerce development.

To understand the indirect drivers, we need to investigate the changes in the overall economic conditions in Hungary. Net monthly earnings, inflation and household consumption are the key influencers.

According to the latest release of the Hungarian Central Statistical Office, the volume of gross domestic product was 2.2% higher in Hungary in the 3rd quarter of 2016 than in the same period of 2015. The actual final consumption of households went up by 3.8%. Household final consumption expenditure was increased by 4.5% (KSH 2016a).

Both raw and calendar-adjusted data show that the volume of sales in retail shops went up by 5.1% compared to the same period of 2015. The volume of sales was increased by 3.9% in specialized and non-specialized food shops, by 5.4% in non-food retail shops and by 7.7% in automotive fuel retailing. Calendar adjusted data show that the volume of sales went up by 4.9% in the period of January–September 2016, compared to the corresponding period of 2015 (KSH 2016b).

According to the latest report by the Hungarian Central Statistical Office (KSH) published in Hungary Today (2016), from January to July in 2016, wages in real terms became significantly higher year-on-year, having gained 7.4%. Average gross and net wages, excluding family tax allowances, reached HUF 257,900 (approx. EUR 833) and HUF 171.500 (approx. EUR 550)

As per BBJ (2016) the indicator for core inflation, excluding indirect tax effects, was 1.4% in September, slightly up from 1.2% in August – according to Hungarian Central Statistical Office (KSH). Households' inflation expectations remained the same at moderate levels, consistent with low underlying inflation developments.

The development level of the digital economy, especially internet and device use must be analyzed as direct drivers of the e-commerce growth.

Both Internet and device use are on the rise in Hungary; especially demand for smartphones and broadband mobile services is becoming stronger. People chiefly use those devices for numerous personal purposes. However, according to the latest data from the European Commission Digital Scoreboard, the use of social networks is the highest in Hungary among the EU countries (EC, 2016a) (see Figure 1). 83% of internet users also use social networks, mainly Facebook. However, internet banking, online shopping and the use of digital public services are the weakest areas in our country, offering the most opportunities for further development.

Note: Social Networks (3b2) is one of the indicators comprised in the Digital Economy and Society Index (DESI). It is part of the Communication (3b) sub-dimension under Use of Internet (3) and represents the percentage of individuals in a country (aged 16-74) who used the Internet to participate in social networks (to create a user profile, post messages, or perform other contributions on Facebook, Twitter, etc.)

The Digital Economy and Society Index (DESI) is an online tool to measure the progress of EU Member States towards a digital economy and society (EC, 2016b). It is a benchmark figure that makes it possible to compare the digital development of EU countries. DESI scores range from 0 to 1; the higher the score, the better the country performance. In 2016, the DESI score for Hungary was 0.47 out of 1.00, which is the 20th rank out of the 28 EU Member States. Figure 2 shows the development of the DESI scores of Hungary and the EU in the last three years. The Digital Economy and Society Index has five principal dimensions: connectivity, human capital, use of internet, integration of digital technology and digital public services.

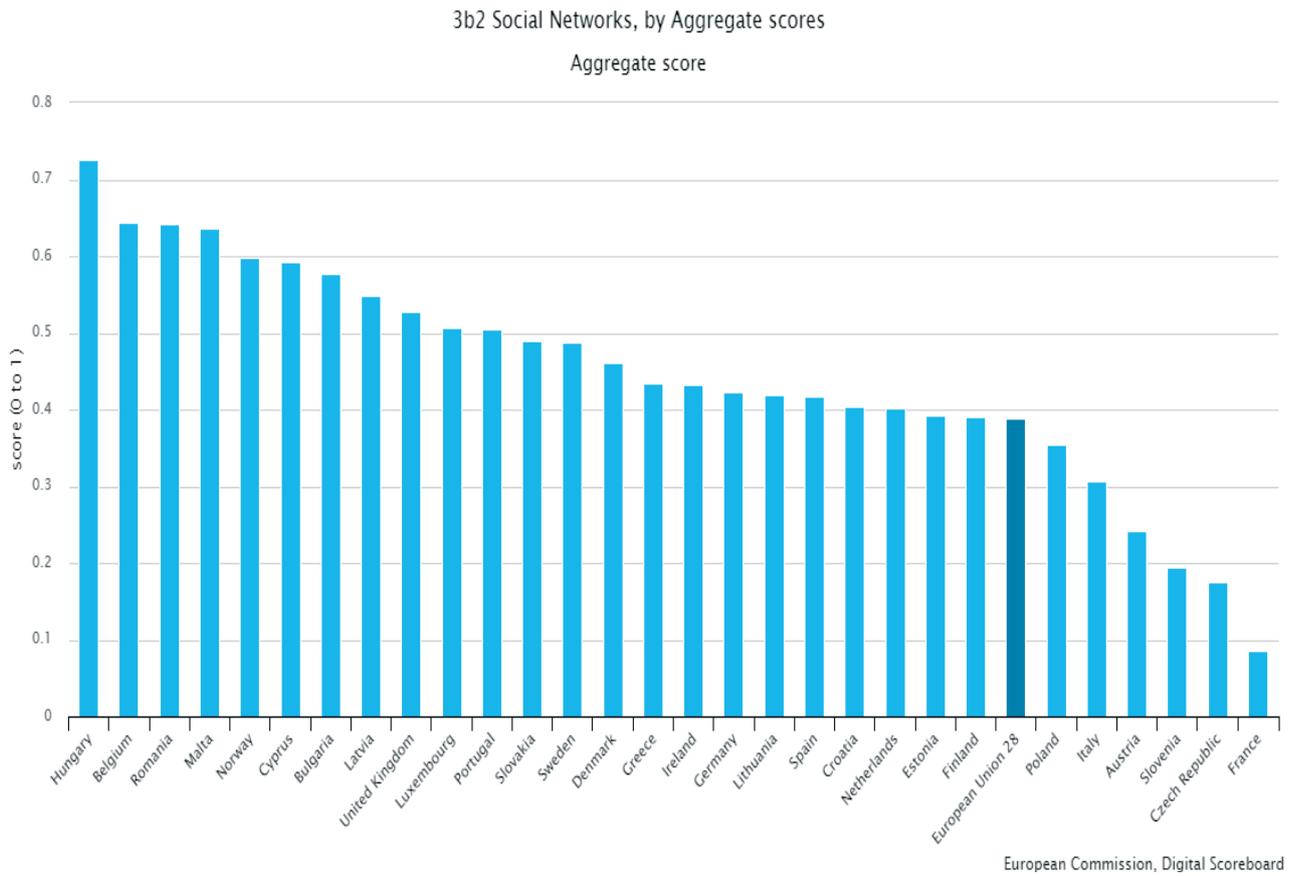

Source: EC (2016a)

*Figure 1. Use of Social Networks in European Union*

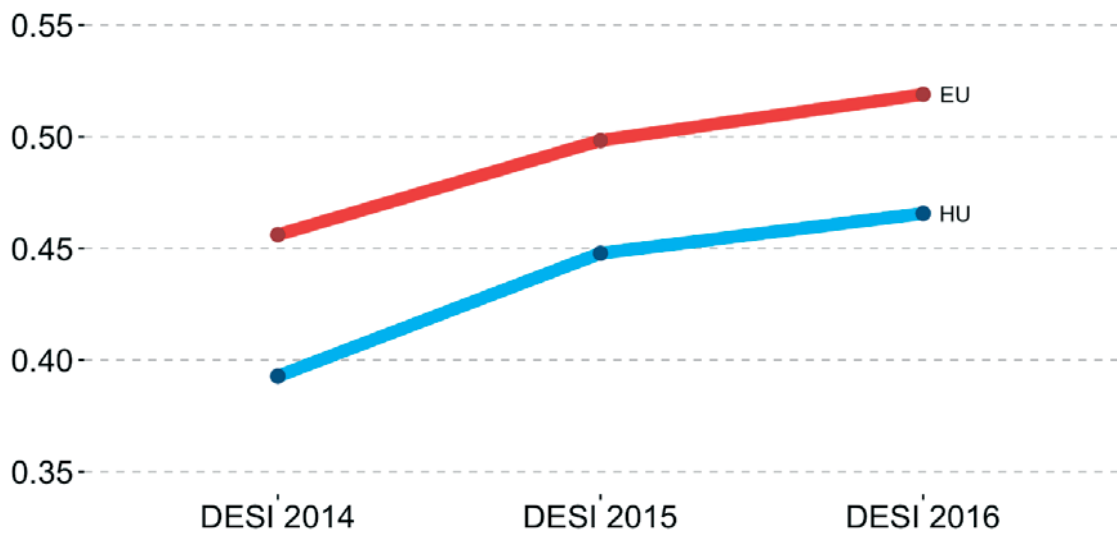

Source: EC (2016b)

*Figure 2. Development of Digital Economy and Society Index (DESI), 2014–2016*

As far as the connectivity component of the DESI Index concerned, there is an advance (0.57 in 2016 from 0.54 in 2015), which is chiefly caused by the progress in the take-up and coverage of fast broadband technologies. As for the use of Internet, the number of Internet users increased to 82% in 2016. Telecommunication plays a fundamental role in the family lives of the Hungarian Internet users, who are convinced that the more telecommunication device a family uses, the better informed it is. According to eNET (2015) Hungarian users are carrying out a variety of activies on the Internet, and outstrip the EU average on the use of the internet. Hungarians think that the Internet makes it easier to get along in life. The most popular device in Hungary is still the computer, but the popularity of smart phones and tablets is also rising, and the use of multiple screens (two or three) is on the rise in Hungary (Consumer Barometer 2016).

The Digital Scoreboard (EC 2016b) reported that, as far as human capital concerned, Hungary shows a diverse picture in digital skills, since only half of the population has at least basic digital skills (compared to the EU average of 55%), whereas ICT specialists represent a relatively high share of the labor force (4.9% compared to 3.7% in the EU). This study also reveals that integration of digital technology by businesses is the biggest problem in Hungary. Hungarian firms should better exploit the possibilities offered by online business, social media and cloud-based applications. This analysis draws attention to the fact that digital public services are also an urgent problem to be solved in Hungary. This field is where Hungary performs fourth from the bottom in the EU. The percentage of e-governance users among the internet users in 2016 was only 32%. This is exactly the same percentage as in 2015.

## DATA AND METHODS

To compile a comprehensive analysis on the current state of e-commerce in Hungary, I conducted external secondary research. I collected and analyzed up-to-date secondary data only from reliable sources such as the European Commission Digital Scoreboard, with special regards to the Digital Scoreboard measuring progress of the European digital economy. Moreover, I collected the latest related reports and data published by the Hungarian Central Statistical Office and Eurostat as well as Google's Consumer Barometer, which is based on the 'The Connected Consumer Survey' in 2016 with more than 1,000 total respondents. Moreover, I used the latest eNET Internet Research and Consulting news and reports, supplemented with the latest related reports by GKI Digital and Budapest Business Journal. My research intention was to blend and analyze up-to-date secondary data published in different reliable sources in a single study to provide deeper insight into this fast-changing industry.

## RESULTS

As a result of the direct and indirect influences, e-commerce volume in Hungary surpassed 131 billion Hungarian forints (about EUR 424 million) in the 2nd quarter of 2016, indicating a 15 billion HUF plus and 18% growth compared to the same period in 2015, according to the latest report by GKI Digital (2016a). The majority of this growth (11 billion HUF) was generated by consumer electronics. This 46% growth compared to Q2 2015 is mainly attributed to the sales promoting effects of the European Football Championship.

The latest report by eNET (2016) revealed that the net online trade volume increased by 17%, while the share of the online retail sector within the overall retail trade went up by 0.4 percentage points in 2015. Moreover, the average shopping basket size increased as well; reaching a record high HUF 11,400 (~EUR 36.6) in 2015, compared to HUF 10,000 (~EUR 32.4) in 2014.

From the European perspective, the current growth rate (18%) of the online retail sector in Hungary is in sync with recent development of the European e-tail sector. The latest EU data by Ecommerce News (2016) show 18.4% growth in 2014 and 18.6% in 2015 respectively.

Share of online retail of the overall trade in Hungary hit 4.1% in 2015 for the first time in history (eNET, 2016). However, as Ecommerce News (2016) shows this ratio is still low compared to the leading markets (see Figure 3), i.e. the UK (15.2%) and Germany (11.6%) or the European average (8.4%).

In 2015, 77% of 6,000 webshops operating in Hungary increased their revenues; only 4% of them reported decrease in revenues. 85% of Hungarian webshops served the Hungarian market only (Piac és Profit, 2016).

According to GKI Digital (2016a), the average conversion rate in Hungary is 2.1%. This means 2.1 purchases per 100 visitors. The best conversion rate (2.6%) can be found in the IT sector, while at the other extreme, in the home and garden sector, the average conversion rate is only 1.4%.

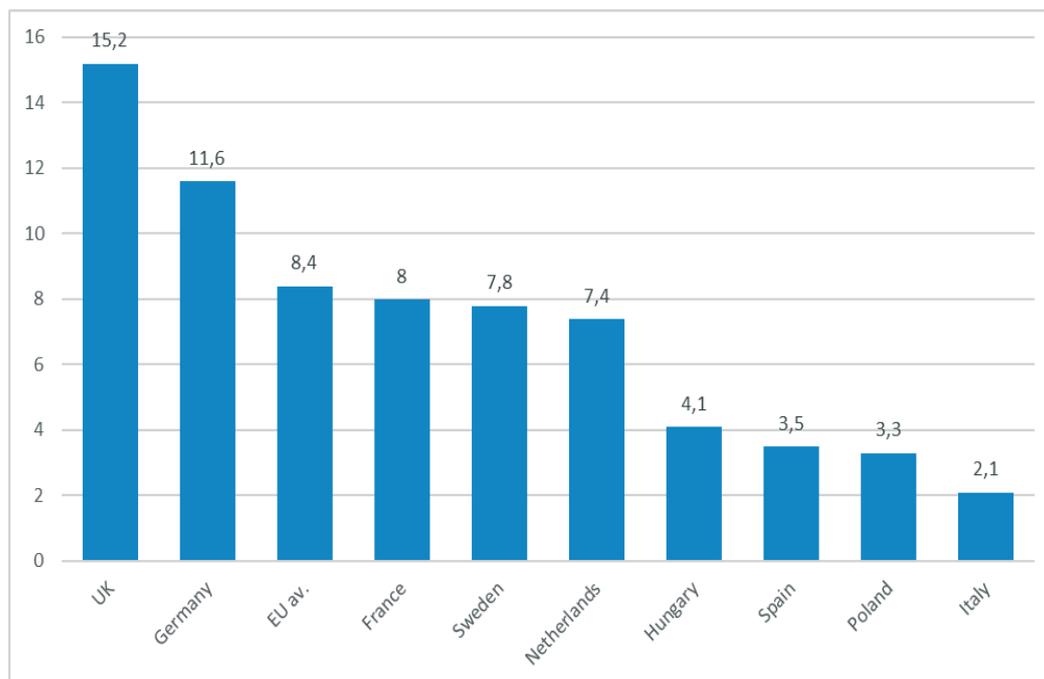

Source: Ecommerce News (2016) and eNet (2016)

*Figure 3. The share of online retail in overall trade (2015)*

GfK (2015) shows a significant concentration of e-commerce in Hungary, as 30 biggest online shops in the industry account for 47% of all orders, while the remaining 53% of orders are distributed among several thousand e-tailers. According to GKI Digital (2016b), the latest report, the top 10 biggest e-tailers account for 1/5 of total orders. The weight of the top 10 significantly increased in 2015, with their aggregated revenue reaching HUF 92 billion HUF (over EUR 300 million). Hence the industry concentration index (C10)=0.34, indicating a very concentrated industry, with 34% aggregated market share of the ten biggest companies in the e-commerce sector in terms of turnover. With 4.7 million orders (21% of the industry total), the top10 e-tailers managed to realize 70% growth rate, while the average growth rate in the industry was "only" 16%.

This report also revealed that the number one e-tailer in terms of sales volume in Hungary is Extreme Digital, followed by two other web-plazas: eMAG and MALL. In this e-commerce giants ranking list, Media Markt (consumer electronics) and Tesco are in the fourth and fifth places. Libri-Bookline (culture) ranked sixth, followed by iPOn (IT) and Aqua (IT). Bonusz Brigad (coupons) and Tchibo (premium lifestyle) are ranked ninth and tenth, respectively.

Most (78%) e-tailers are active in social media (mainly on Facebook). For most webshops promotion means offering discounts. The size of the average discount in the industry is 12% (GKI Digital, 2016a).

As far as the online shoppers concerned, the percentage of individuals using the internet for ordering goods or services rose 36%, which is still below the European average (53%) (Figure 4).

The largest (more than 10 percentage point) increases in the percentage of individuals using the internet for purchasing between 2012 and 2016 were recorded in Lithuania, the Czech Republic, Ireland, Hungary, Spain, Italy and Slovakia (Eurostat, 2016).

According to an infographic on e-commerce in Hungary (eNET 2015), the profile of the typical online shopper can be described as follows: married woman, aged 18-39, without children, living in a city and shopping online for more than 3 years. The typical online shopper purchases online at least once every three months.

eNET (2015) also revealed the most important online shopping motivations. A favorable price is in the first place (50%), followed by home delivery (48%) and the ability to compare prices (45%). It is also a motive that something is available online only (41%), or that a wide variety of products can be bought online (40%). Hungarian research findings by eNET (2015) have partially supported the study by Chaparro-Peláez, et al. (2016), as one factor of convenience, home delivery proved to be one of the most important motives of online shopping in Hungary. However, eNET (2015) findings are definitely contrary to the conclusion reached by Chaparro-Peláez et al. (2016) that price is not a motivation anymore for consumers in online shopping. In Hungary, (good) price is a decisive factor, the strongest motivation to buy online.

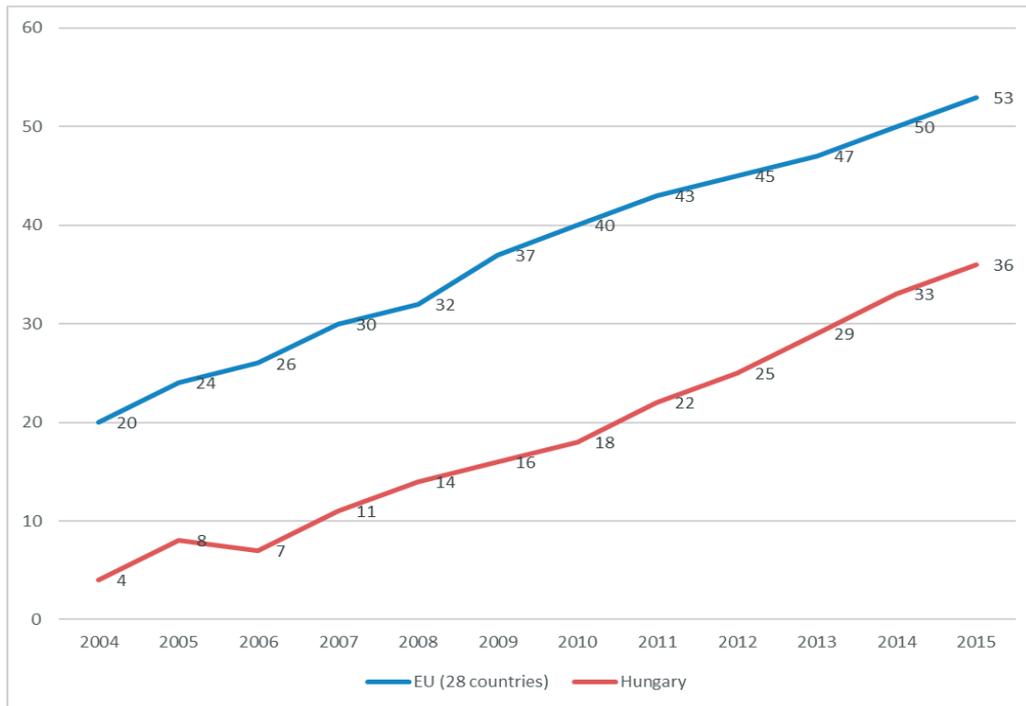

Source: Eurostat (2015), Note: % of individuals aged 16 to 74

*Figure 4. The percentage of individuals using the internet for purchasing (2015)*

The distribution of online shoppers across the seven regions in Hungary is uneven. The majority (36%) of them live in Central Hungary (the Budapest region). On the contrary, in other regions this ratio is only between 9% and 13% (eNET 2015).

The majority of Hungarian online shoppers prefer Hungarian webshops. Hungarian-based online retailers account for 88% of total online orders, while the share of foreign webshops is only 12%. However, in 2015, the number of Hungarians buying goods from abroad rose by 14%. Hungarian shoppers' favorite e-commerce sites are eBay, Alibaba, Booking.com and Amazon.com (GfK, 2015). People who ordered items from abroad mostly purchased clothing and sportswear, as well as IT devices, watches and jewelry from foreign webshops.

As far as payment methods concerned, webshops usually offer cash payment upon delivery, bank transfer, and cash payment in a shop, site or office. Cash payment is still the most popular payment method amongst online shoppers, i.e. cash still rules in Hungary (eNET, 2016).

As to delivery methods, the same report reveals that courier services including postal services are the favorite options. This delivery method is available in 90% of e-tailers. Courier service is followed by personal collection in a shop, and personal collection at staffed stores operated by partners. The most preferred delivery method based on the total value of the transactions was delivery by courier services, with a 42% share.

As far as industry trends concerned, even stronger marketing activities are the key to success (eNET, 2015). The best industry players are using price comparison webpages, SEO, paid online advertisements, newsletters and Facebook ads. Moreover, there is a need for wider selection, and for expanding warehouse capacity as well as innovation of services and webpages.

# CONCLUSIONS AND OUTLOOK

As research findings show, e-commerce in Hungary is in a boom phase. Therefore, it is strongly recommended for traditional brick-and-mortar companies that are currently present in the Hungarian market to expand their operation into a brick-and-click company. Pure-click companies are also welcomed in this emerging e-commerce market, which is now becoming stronger and stronger, with online retail expected to grow further. Growth is not expected to cease in the coming years; there is more growth potential, since the current ratio of e-commerce (4.1%) is less than half the European average (8.4%). Direct and indirect drivers of e-commerce in Hungary provide favorable conditions for further growth.

There are signs of further concentration of the e-commerce industry in Hungary. Big players in the e-commerce sector will reach even more customers, while small players will become even more specialized. The good news is that the size of the cake is expected to grow further, as more and more Hungarians become interested in shopping online. However, the ratio of foreign webshops being used by Hungarian customers is expected to increase in the future. Therefore, competition in e-commerce will be fierce in the long run. Experts also predict that the share of purchases initiated with mobile devices (m-commerce) will go up, so companies must be prepared for this.

To sum-up, the net e-commerce sales volume will probably continue to increase in the next 3–4 years, with no major hurdles currently in sight. Considering e-commerce trends in Western European countries, there is still plenty of potential for growth and progress in Hungary.

The limitation of this study is the use of secondary information sources only, which narrows down the scope of applied research techniques. As far as further direction of research concerned, it would be advisable to conduct a representative, nationwide survey to explore the explicit and latent relationships of decisive factors of e-commerce and to segment online shoppers in Hungary.

REFERENCES

ANVARI, R. D. & NOROUZI D. (2016). The Impact of E-commerce and R&D on Economic Development in Some Selected Countries, Procedia - Social and Behavioral Sciences. Volume 229. 354–362. http://dx.doi.org/10.1016/j.sbspro.2016.07.146

APĂVĂLOAIE, E.-I (2014). The impact of the internet on the business environment. Procedia Economics and Finance. 15 (2014), 951–958, http://dx.doi.org/10.1016/S2212-5671(14)00654-6

BBJ (2016). MNB measures of underlying inflation up in September. Budapest Business Journal. Retrieved: 11 October, 2016. http://bbj.hu/economy/mnb-measures-of-underlying-inflation-up-in-september_123205

CHAPARRO-PELÁEZ, J., AGUDO-PEREGRINA Á. F. & PASCUAL-MIGUEL F.J. (2016). Conjoint analysis of drivers and inhibitors of e-commerce adoption. Journal of Business Research. Volume 69. Issue 4. April 2016. 1277–1282. http://dx.doi.org/10.1016/j.jbusres.2015.10.092

CHOSHIN, M. & GHAFFARI, A. (2017) An investigation of the impact of effective factors on the success of e-commerce in small- and medium-sized companies. Computers in Human Behavior. Volume 66. 67–74. http://dx.doi.org/10.1016/j.chb.2016.09.026

CONSUMER BAROMETER (2016) Which devices do people use? Retrieved: 16 December, 2016. https://www.consumerbarometer.com/en/graph-builder/?question=M1&filter=country:hungary

EC (2016a) European Commission. Digital Scoreboard. Retrieved: 16 December, 2016 https://ec.europa.eu/digital-single-market/en/social-networks-desi-indicator-3b2

EC (2016b) European Commission. Digital Economy and Society Index (DESI). Hungary Profile. Retrieved: 16 December 2016. https://ec.europa.eu/digital-single-market/en/scoreboard/hungary

ECOMMERCE NEWS (2016). Ecommerce in Europe. Retrieved: 16 September, 2016. http://ecommercenews.eu/ecommerce-per-country/ecommerce-in-europe/

ENET (2015). E-kereskedelmi Körkép 2015. (E-commerce Report) Retrieved: 28 December, 2015. http://www.enet.hu/hirek/e-kereskedelmi-korkep-2015/?lang=hu

ENET (2016). Hungarians spend one in HUF 25 online. Retrieved: 13 June, 2016. http://www.enet.hu/news/hungarians-spend-one-in-huf-25-online/?lang=en,

EUROSTAT (2015) Individuals using the internet for ordering goods or services. Retrieved: 12 September 2015. http://ec.europa.eu/eurostat/tgm/table.do?tab=table&init=1&language=en&pcode=tin00096&plugin=1

EUROSTAT (2016) E-commerce statistics for individuals. About two thirds of internet users in the EU shopped online in 2016. Retrieved: 22 December, 2016. http://ec.europa.eu/eurostat/statistics-explained/index.php/E-commerce_statistics_for_individuals

FALK, M. & HAGSTEN, E. (2015). E-commerce trends and impacts across Europe. International Journal Production Economics. Volume 170, Part A, December 2015, 357–369. http://dx.doi.org/10.1016/j.ijpe.2015.10.003

GABRIELSSON, M. & GABRIELSSON, P. (2015). Internet-based sales channel strategies of born global firms. International Business Review. Volume 20 (1), 88–99. http://dx.doi.org/10.1016/j.ibusrev.2010.05.001

GfK (2015). A magyar online vásárlók körében kedveltek a külföldi webshopok. (Hungarians like foreign webshops) Retrieved: 04 June, 2015. http://www.gfk.com/hu/insightok/press-release/a-magyar-online-vasarlok-koereben-kedveltek-a-kuelfoeldi-webshopok

GKI Digital (2016a) Dübörög az online szektor. (Online sector is roaring) Retrieved: 01 September 2016. http://www.gkidigital.hu/2016/09/01/duborog-az-online-szektor/

GKI Digital (2016b). E-toplista 2016 – A legnagyobb webáruházak listája. (E-tail Top List 2016 - List of the biggest webshops.) Retrieved: 2 June, 2016. http://www.gkidigital.hu/2016/06/02/etoplista2016/

HUNGARY TODAY (2016) Stats Office: Average Monthly Net Wage In Hungary Continues To Grow To €550 Retrieved: 20 September, 2016. http://hungarytoday.hu/news/stats-office-avarage-monthly-net-wage-hungary-continues-grow-e550

JOHANSSON, T. & KASK, J. (2017). Configurations of business strategy and marketing channels for e-commerce and traditional retail formats: A Qualitative Comparison Analysis (QCA) in sporting goods retailing. Journal of Retailing and Consumer Services. Volume 34. 326–333. http://dx.doi.org/10.1016/j.jretconser.2016.07.009

KSH (2016a) Gross domestic product (GDP), 3rd quarter 2016. Hungarian Central Statistical Office. Retrieved: 6 December 2016. https://www.ksh.hu/docs/eng/xftp/gyor/gdp/egdp1609.html

KSH (2016b) A 5.1% increase in the sales of retail shops. Hungarian Central Statistical Office. Retrieved: 7 November 2016. http://www.ksh.hu/docs/eng/xftp/gyor/kie/ekie1609.html

PANTANO E., PRIPORAS C.-V., SORACE, S. & IAZZOLINO, G. (2017) Does innovation-orientation lead to retail industry growth? Empirical evidence from patent analysis. Journal of Retailing and Consumer Services. Volume 34. 88–94. http://dx.doi.org/10.1016/j.jretconser.2016.10.001

PIAC ÉS PROFIT (2016) Nő a webáruházak forgalma, pedig nem lépik át a határt. (Revenue of webshops is increasing, though they do not cross the border) Retrieved: 13 July, 2016. http://www.piacesprofit.hu/kkv_cegblog/no-a-webaruhazak-forgalma-pedig-nem-lepik-at-a-hatart

STATISTA (2016). E-commerce in Hungary. Retrieved: 18 January, 2017 https://www.statista.com/outlook/243/139/e-commerce/hungary